\documentclass[aps,epsfig,prb,unsortedaddress,twocolumn,longbibliography]{revtex4-1}

\usepackage{amsmath}
\usepackage{graphicx}

\def\nn{\nonumber}

\begin{document}
\title{Theory of a Carbon-Nanotube Polarization Switch}

\author{Ken-ichi Sasaki}
\email{sasaki.kenichi@lab.ntt.co.jp}
\affiliation{NTT Basic Research Laboratories, NTT Corporation,
3-1 Morinosato Wakamiya, Atsugi, Kanagawa 243-0198, Japan}

\author{Yasuhiro Tokura}
\affiliation{Faculty of Pure and Applied Sciences,
University of Tsukuba, Tsukuba, Ibaraki 305-8571, Japan}



\date{\today}

\begin{abstract}
 Recently, it was suggested that the polarization dependence of light
 absorption to a single-walled carbon nanotube is altered by carrier
 doping.
 We specify theoretically the doping level at which the polarization
 anisotropy is reversed by plasmon excitation.
 The plasmon energy is mainly determined by the diameter of a nanotube,
 because pseudospin makes the energy independent of the details of the
 band structure.
 We find that the effect of doping on the Coulomb interaction appears
 through the screened exchange energy, which can be observed as changes
 in the absorption peak positions.
 Our results strongly suggest the possibility that oriented nanotubes
 function as a polarization switch.
\end{abstract}

\maketitle

\section{Introduction}

\begin{figure}[htbp]
 \begin{center}
  \includegraphics[scale=0.32]{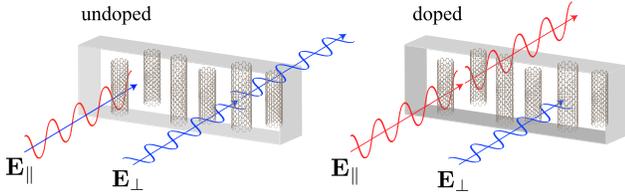}
 \end{center}
 \caption{{\bf Optical anisotropy of oriented CNTs.} 
 Oriented and undoped (doped) CNTs pass ${\bf E}_\perp$ (${\bf
 E}_\parallel$) only. 
 Thus, oriented CNTs function as a polarization switch.
 }
 \label{fig:cntpolarizer}
\end{figure}

A carbon nanotube (CNT)~\cite{Iijima1993}
absorbs light whose linear polarization is parallel to the tube's axis
(${\bf E}_\parallel$), 
but not when the polarization is perpendicular to it (${\bf E}_\perp$).~\cite{Ajiki1994,Hwang2000,ichida04}
The optical anisotropy of a CNT enables oriented
CNTs to function as an optical polarizer.~\cite{Shoji2008,Kang2010}
Recently, it was theoretically predicted that 
the polarization dependence is reversed by charge doping;~\cite{Sasaki2016a}
a doped CNT transmits ${\bf E}_\parallel$ and
absorbs ${\bf E}_\perp$ (see Fig.~\ref{fig:cntpolarizer}).

The absorption of ${\bf E}_\perp$ originates from the
resonant excitation of collective oscillations of electrons (plasmon),
which differs entirely from the excitation of individual electrons or
excitons by ${\bf E}_\parallel$ in an undoped CNT.~\cite{Abajo2014,Sasaki2016a}
This theory of plasmon resonance accounts qualitatively for the
anomalous absorption peaks observed experimentally in doped CNTs.~\cite{Kazaoui1999,Liu2003,Kramberger2012,Igarashi2015} 
However, because the theoretical conclusion was derived using the Drude
model which only takes account of intra-band
electronic transitions,
the exact doping and chirality~\cite{saito98book} 
dependencies of the absorption spectrum remain unknown.

In this paper we elucidate these dependencies by investigating the competition
between intra and inter-band transitions with the Kubo formula.
On the basis of predicted doping and chirality dependencies,
we conclude that doped CNTs absorb ${\bf E}_\perp$
over frequencies ranging from infrared to visible.
This expands the application range of CNT polarizers
and suggests the possibility that the polarization direction of
transmitted light is changed by 90 degrees with doping rather than by
spatial rotation.

This paper is organized as follows.
In Sec.~\ref{sec:selection},
we explain the optical selection rule of CNTs.
By calculating the dynamical conductivity, we show that momentum
conservation and pseudospin play very essential role in determining the
possible transitions.
In Sec.~\ref{sec:plasmon}, we examine absorption spectra for armchair
and zigzag CNTs, which are the main result of this paper.
The effect of Coulomb interaction 
on the absorption spectra is studied in Sec.~\ref{sec:exciton}.
Our discussion is provided in Sec.~\ref{sec:discussion}.
The calculation details which are necessary to reproduce the results
of Secs.~\ref{sec:selection} and~\ref{sec:plasmon}
are given in Appendix.

\section{Selection Rule}\label{sec:selection}

\subsection{Parallel Polarization}

\begin{figure}[htbp]
 \begin{center}
  \includegraphics[scale=0.39]{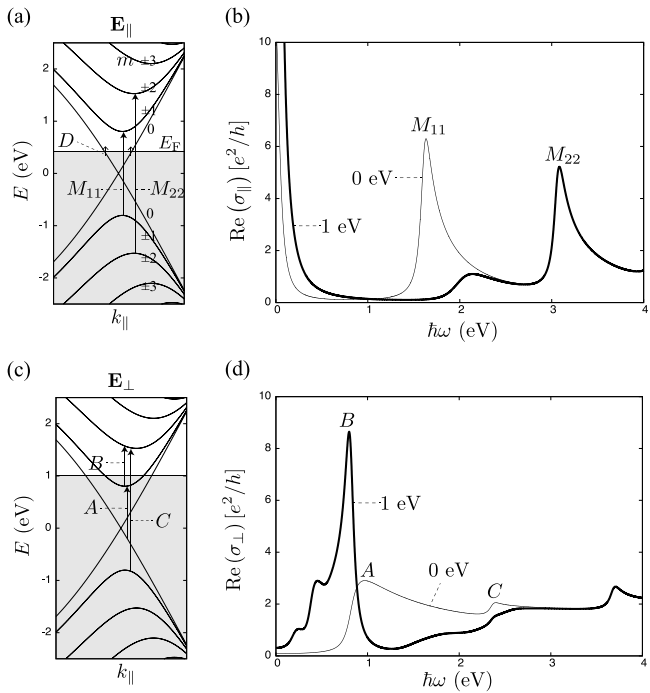}
 \end{center}
 \caption{
 {\bf Selection rule of armchair CNTs.} 
 (a) The direct transitions caused by ${\bf E}_\parallel$ ($\Delta m=0$) are denoted by arrows.
 (b) Calculated real part of the dynamical conductivity along
 ${\bf E}_\parallel$, ${\rm Re}(\sigma_\parallel)$, is shown for
 different Fermi energy positions.
 The vertical axis is given in units of $e^2/h$, where $e$ is electron charge magnitude and $h$ Planck constant.
 (c) The momentum selection rule of ${\bf E}_\perp$ is
 $\Delta m=\pm1$.
 The inter-band transitions with $m=0 \to \pm1$ are denoted by $A$ and
 these are allowed (forbidden) for low (high) doping. 
 When $E_F=1$ eV, the intra-band transitions with
 $m=+1\to +2$ or $-1\to -2$ are allowed and these are denoted by $B$.
 The transition $C$ is suppressed by pseudospin.
 (d) ${\rm Re}(\sigma_\perp)$ is shown for different Fermi energy positions.
 }
 \label{fig:Absorption_armchair}
\end{figure}

The electronic transition caused by ${\bf E}_\parallel$
is a direct transition without a change in momentum of a photo-excited electron.~\cite{Ajiki1994}
In the band-diagram of a $(10,10)$ armchair CNT shown in
Fig.~\ref{fig:Absorption_armchair}(a), each of 
the band curves plotted as a function of the wavevector along the
tube's axis ($k_\parallel$) is an eigenstate of the
momentum around it and specified by magnetic quantum number $m$.~\cite{Minot2004}
The two bands with linear dispersion that cross each other at $E=0$ have vanishing
$m$, while the other curve is degenerate ($\pm|m|$) 
corresponding to the clockwise and anticlockwise circumferential
motions, 
and the magnitude $|m|$ increases with the energy $|E|$.
Because the same $m$ value appears in the conduction and valence bands
symmetrically with respect to $E=0$,
there are two possible cases of direct transition:
transition between the valence and conduction bands 
(inter-band transition) or within either band 
(intra-band transition).

The doping dependence of the direct inter-band
transition is roughly known from Fermi-Dirac statistics.
When the doping level is low, e.g. $E_F=0$ eV (undoped),
the direct inter-band transitions denoted by 
$M_{11}$ and $M_{22}$ in
Fig.~\ref{fig:Absorption_armchair}(a) 
are both allowed by the Pauli exclusion principle,~\cite{Petit1999a,Malic2006}
while when the doping level is high, e.g. $E_F=1$ eV, 
$M_{11}$ is forbidden, although $M_{22}$ is still allowed.
In Fig.~\ref{fig:Absorption_armchair}(b), the calculated real part of
the dynamical conductivity ${\rm Re}(\sigma_\parallel)$ shows that
the $M_{11}$ peak disappears when $E_F=1$ eV.
Meanwhile, a Drude peak corresponding to the direct intra-band
transition denoted by $D$ in
Fig.~\ref{fig:Absorption_armchair}(a) 
develops in the zero-frequency limit of
${\rm Re}(\sigma_\parallel)$. 
The peak intensity increases with doping
because the density of states at $E_F$ increases with doping. 
The disappearance of the $M_{11}$ peak
and enhancement of the Drude peak are evidence of high doping that is
provided by the absorption spectra of ${\bf E}_\parallel$.

\subsection{Perpendicular Polarization}

The electronic transition caused by ${\bf E}_\perp$ is the indirect
transition, and transitions with an $m$ change of $\pm1$, $\Delta m=\pm1$, are
dominant over transitions with $|\Delta m|\ge 2$.
This selection rule is a consequence of momentum conservation being
applied to a case where, at the surface of a CNT, the azimuthal component of 
${\bf E}_\perp$ is approximately written as a sine (or cosine) function
of the azimuthal angle ($\theta$) of the cylinder.~\cite{Ajiki1994}
More exactly, 
this is a consequence of momentum conservation being used in combination
with the two facts that a plane wave is a superposition of different
magnetic quantum numbers and that tube diameter $d_t$ of nanometer scale
is much shorter than the light wavelength of micrometer scale.~\cite{Bohren1983,Sasaki2016a}
The proof goes as follows.
When the light polarization is set perpendicular (${\bf e}_x$) to a
tube's axis (${\bf e}_z$), an incident plane wave of frequency $\omega$
and amplitude $E_{\rm in}$ is written as
$E_{\rm in} e^{i(ky-\omega t)} {\bf e}_x$. 
In a cylindrical coordinate system $(r,\theta,z)$, 
the field is expressed as
\begin{align}
 {\bf E}_\perp(r,\theta;t)
 =E_{\rm in}
 e^{i(kr\sin\theta -\omega t)}
 \begin{pmatrix}
  \cos\theta \cr
  -\sin\theta \cr
  0
 \end{pmatrix}.
\end{align}
By using the formula for the Bessel functions,
$e^{ikr\sin\theta} =\sum_{m=-\infty}^{\infty} J_m(kr) e^{im\theta}$, 
we obtain the azimuthal component as
\begin{align}
 E_{\theta} = \sum_{m}\frac{E_{\rm in}}{2i} 
 \left( J_{m+1}(kr)-J_{m-1}(kr) \right) e^{i(m\theta-\omega t)}.
\end{align}
Because $J_n(kr)\propto (kr)^n$,
$|E_{\theta}|$ is dominated by the modes with $m=\pm1$ when $kr \ll 1$, 
which we assume throughout this paper.~\footnote{
It is possible that in electron energy loss spectroscopy,
de Broglie wavelength may be comparable to $d_t$ and 
electronic transitions satisfying $|\Delta m|\ge 2$ are allowed.
Electronic transitions with $|\Delta m|= 2$ were demonstrated
via Raman spectroscopy for an undoped CNT placed in a metal
nanogap,~\cite{Takase2013} where the light wavelength is reduced
by the plasmon at the nanogap.
}
Applying the momentum selection rule to the band-diagram in 
Fig.~\ref{fig:Absorption_armchair}(c) 
we can expect that the transitions denoted by $A$
and $B$ to develop the peaks in ${\rm Re}(\sigma_\perp)$ for an
undoped and doped CNT, respectively, and these are confirmed in
Fig.~\ref{fig:Absorption_armchair}(d).

\begin{figure}[htbp]
 \begin{center}
  \includegraphics[scale=0.4]{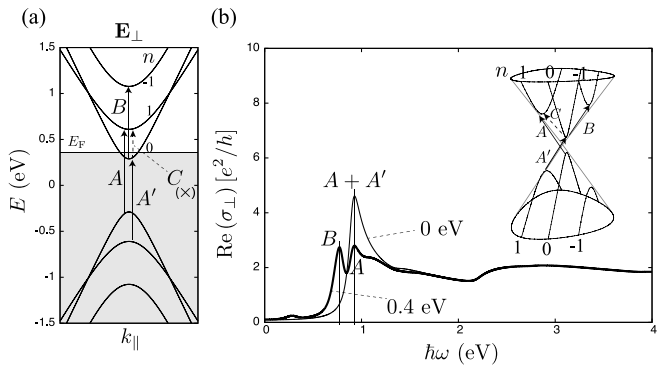}
 \end{center}
 \caption{
 {\bf Selection rule of zigzag CNTs.} 
 (a) The inter-band transitions denoted by $A$ and $A'$ are allowed 
 when $E_F=0$ eV.
 Because $A$ and $A'$ are the forward scattering,
 a single peak develops in ${\rm Re}(\sigma_\perp)$ as shown in (b).
 When $E_F=0.4$ eV, the peak intensity halves because $A'$ is not
 allowed by exclusion principle whereas $A$ is still allowed.
 Besides, the intra-band transition $B$ is allowed by pseudospin.
 The three dimensional band-diagram shows that the transitions $A$,
 $A'$, and $B$ are all the forward scattering, while $C$ is the backward
 scattering that is forbidden by pseudospin.
 }
 \label{fig:Absorption_szig}
\end{figure}

The selection rule, $\Delta m=\pm1$, explained above is a result of the
momentum conservation only, and the indirect transitions are further
restricted to the forward scattering by the symmetry that originates
from the two sublattices nature of the electronic wavefunction known as
pseudospin.~\cite{Sasaki2011}
For example, it suppresses a transition (of the backward scattering) denoted
by $C$ in Fig.~\ref{fig:Absorption_armchair}(c) to
develop a strong peak in ${\rm Re}(\sigma_\perp)$ like
the $M_{11}$ and $M_{22}$ peaks although it is a transition between band-edges
with a large density of states.
A profound effect of pseudospin on 
the selection rule is more clearly seen for a doped semiconducting CNT.
In the band-diagram of a $(16,0)$ zigzag CNT shown in
Fig.~\ref{fig:Absorption_szig}(a), each of the band curves is specified
by shifted magnetic quantum number $m=\pm m_0 +
n$ where $m_0$ is a nonzero integer ($m_0=11$) and $n=0,\pm 1,\cdots$.
When $E_F=0.4$ eV, the transition denoted by $C$ with $\Delta m=1$ is
allowed by the momentum selection rule, however, it is actually forbidden by pseudospin.
Meanwhile the transition denoted by $B$ with $\Delta m=-1$ is fully allowed.
This difference is peculiar because the transition energy of $C$ is
smaller than that of $B$.
It becomes clear that
$B$ ($C$) is forward (backward) scattering by drawing the
three-dimensional band-diagram in the inset of
Fig.~\ref{fig:Absorption_szig}(b).
As a result of pseudospin the peak position in 
${\rm Re}(\sigma_\perp)$ is approximately given by 0.8 eV.
The peak position of the doped zigzag CNT is similar to that of the
doped armchair CNT ($B$ in Fig.~\ref{fig:Absorption_armchair}(d))
regardless of the difference of the band-diagrams of the zigzag and
armchair CNTs.
It should be noted that 
the lack of the transition $C$ with $\Delta m=1$ 
does not mean that there is an asymmetry between 
the clockwise and anticlockwise circumferential motions of the electrons.
Each band with the index $n$ is actually degenerate ($\pm m_0$) 
corresponding to the different valleys, and the subband with $n$ in one 
valley relates with the subband with $-n$ in the other valley.
Thus the lack of the transitions with $\Delta m=1$ in one valley means 
the transitions with $\Delta m=1$ in different valley are allowed.

\section{Depolarization and plasmon}\label{sec:plasmon}

\begin{figure}[htbp]
 \begin{center}
  \includegraphics[scale=0.3]{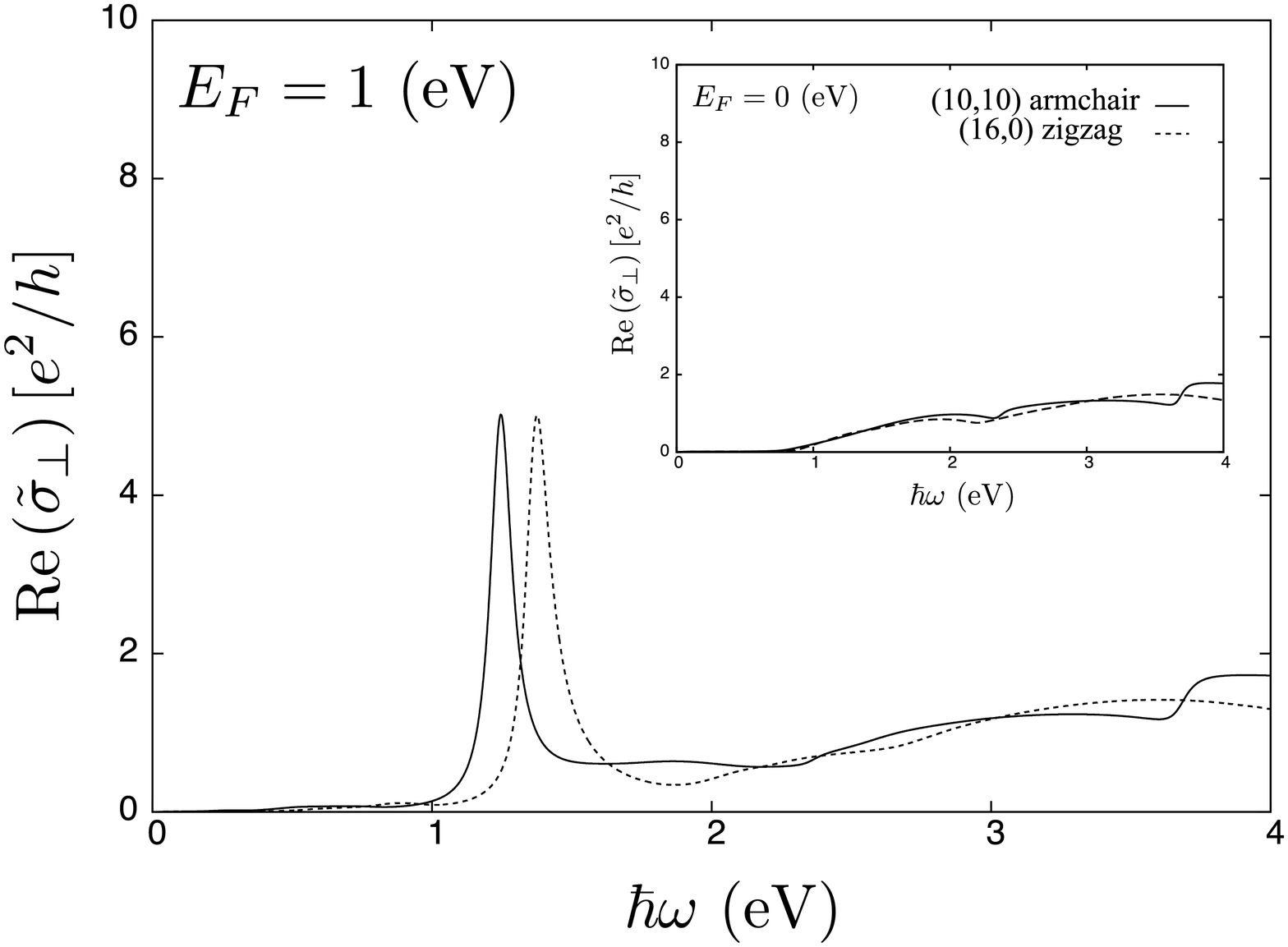}
 \end{center}
 \caption{
 {\bf Calculated absorption spectra for doped armchair and zigzag CNTs.} 
 The absorption spectrum that includes the depolarization effect is
 given by the real part of the dynamical conductivity divided by the
 relative permittivity 
 $\varepsilon_\perp$, ${\rm Re}(\sigma_\perp/\varepsilon_\perp)$
 ($\equiv {\rm Re}(\tilde{\sigma}_\perp)$).
 The peak originates from a resonant excitation of plasmon caused by doping.
 The inset shows the depolarization effect in undoped CNTs. 
 }
 \label{fig:plasmonpeak}
\end{figure}

According to the selection rule only, 
we may expect the peaks caused by ${\bf E}_\perp$,
such as $A$ in Fig.~\ref{fig:Absorption_armchair}(d) and $A+A'$ in Fig.~\ref{fig:Absorption_szig}(b),
to appear in the absorption spectra of undoped CNTs.~\cite{Bozovic2000}
However, it is not.
The calculated absorption spectrum which is given by $\sigma_\perp$
divided by the relative permittivity $\varepsilon_\perp$ as 
${\rm Re}(\sigma_\perp/\varepsilon_\perp)$ ($\equiv {\rm Re}(\tilde{\sigma}_\perp)$)
does not exhibit the corresponding peak when $E_F=0$ eV, 
as shown in the inset of Fig.~\ref{fig:plasmonpeak}.
This is widely known as the depolarization effect.~\cite{Ajiki1994,Hwang2000,ichida04} 
As a result of the momentum transfer from ${\bf E}_\perp$ to an electron, 
a non-uniform density distribution around the tube's axis similar to an
electric dipole is introduced and induces a depolarization field.~\cite{Ajiki1994}
When the doping level is low, 
the depolarization field almost cancels out the applied field,
and the total field defined by the sum of the applied and depolarization
fields, is suppressed. 
Even though the electronic transition is allowed by the selection rule, 
the electron does not undergo a transition since the electric field
by itself almost disappears due to the depolarization effect.

The main point of this paper is that the efficacy of the depolarization
field depends strongly on doping. 
When the doping level is as high as $E_F = 1$ eV,
absorption peaks develop in the region $\hbar \omega=1.2\approx
1.3$ eV as shown in Fig.~\ref{fig:plasmonpeak}.
It can be shown that these peaks originate from the fact that 
the depolarization field is strongly enhanced at the specific frequency.
Even if an infinitely small electric field is applied to a doped CNT, 
the depolarization field has a finite amplitude. 
This state is produced by the self-sustaining collective motion of the
electrons (plasmon or plasmon-polariton), which is in sharp
contrast to the single-particle excitation constituting the absorption
peaks of ${\bf E}_\parallel$.
Meanwhile, the Drude peak is absent for ${\bf E}_\perp$,
which is also in contrast to the case of ${\bf E}_\parallel$.

\begin{figure}[htbp]
 \begin{center}
  \includegraphics[scale=0.5]{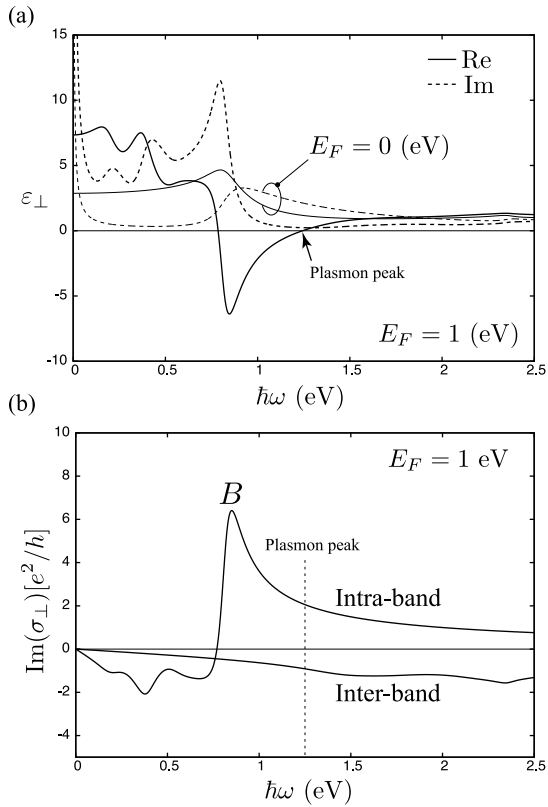}
 \end{center}
 \caption{{\bf Permittivity of (10,10) armchair CNTs.}
 (a) ${\rm Re}(\varepsilon_{\perp})$
 (${\rm Im}(\varepsilon_{\perp})$) when $E_F=1$ eV is shown by the solid
 (dashed) curve.  
 The results when $E_F=0$ eV are shown as a reference.
 (b) 
 The intra-band transitions caused by high doping 
 are essential in the appearance of a plasmon peak.
 Meanwhile the contribution of inter-band transitions is not negligibly small.
 }
 \label{fig:sigma}
\end{figure}

The total electric field that the electrons in a CNT really ``see'' is
given by the applied field divided by the relative permittivity ${\bf E}_\perp/\varepsilon_\perp$.
Mathematically, it is shown that by solving Maxwell equations
while taking account of the boundary conditions at the tube's surface,~\cite{Sasaki2016a}
$\varepsilon_\perp$ is written as
\begin{align}
 \varepsilon_\perp =1 - \frac{\sigma_\perp(E_F)}{i\omega
 \epsilon d_t},
 \label{eq:loopcor}
\end{align}
where $d_t$ is the diameter of a CNT and $\epsilon$ is the permittivity
of the surrounding medium.~\cite{Ajiki1994}
To observe that the vanishing real part of $\varepsilon_\perp$ is
essential for the appearance of plasmon, 
we plot the real and imaginary parts of $\varepsilon_\perp$ as a function of energy in
Fig.~\ref{fig:sigma}(a) for undoped and doped CNTs.
Indeed, when $E_F=1$ eV, ${\rm Re}(\varepsilon_\perp)$ vanishes at an energy
that corresponds to the absorption peak position seen in
Fig.~\ref{fig:plasmonpeak}, where a small magnitude of ${\rm
Im}(\varepsilon_\perp)$ helps
the total electric field to enhance in a resonant fashion.
Note that in the present calculations
a surrounding medium with $\epsilon=2\epsilon_0$ is
assumed~\cite{Igarashi2015} where $\epsilon_0$ is the permittivity of
free space,
and that a large value of $\epsilon/\epsilon_0$ has the advantage of
decreasing the plasmon energy, 
because ${\rm Re}(\varepsilon_\perp)$ shifts upward in effect and zero
of which shows a redshift.

To understand the cause of the appearance of
plasmons in the doped CNTs more clearly, 
we consider the relative significance of the contributions made
by intra and inter-band transitions to ${\rm Re}(\varepsilon_\perp)$.~\cite{Bohren1983}
By noting that in Eq.~(\ref{eq:loopcor})
${\rm Re}(\varepsilon_\perp)$
is proportional to the imaginary part of the dynamical
conductivity, ${\rm Im}(\sigma_\perp)$,
we show each contribution, ${\rm Im}(\sigma^{\rm intra}_\perp)$ and ${\rm Im}(\sigma^{\rm inter}_\perp)$, for the representative case of high
doping level ($E_F=1$ eV) in Fig.~\ref{fig:sigma}(b).
${\rm Im}(\sigma^{\rm inter}_\perp)$ is a negative value for the
frequency range of interest. 
Thus, if we neglect the intra-band transitions, 
${\rm Re}(\varepsilon_\perp) > 0$ and the condition for plasmon existence is unsatisfied. 
When $E_F=1$ eV, 
the contribution made by the intra-band transition ($B$) causes a peak
in ${\rm Im}(\sigma^{\rm intra}_\perp)$ so that 
${\rm Re}(\varepsilon_\perp)$ exhibits a dip at $\hbar \omega \approx 0.8$ eV.
With increasing $\hbar \omega$ from a dip, 
${\rm Im}(\sigma_\perp^{\rm intra})$ decreases and 
$-{\rm Im}(\sigma_\perp^{\rm inter})$ increases.
As a result, ${\rm Re}(\varepsilon_\perp)$ becomes zero at around 1.2 eV
and the sign of ${\rm Re}(\varepsilon_\perp)$ changes at the energy.
It is interesting that when combined with the intra-band transitions,
the contribution of the inter-band transitions to
the dynamical conductivity is not negligible since it tends to redshift
the plasmon energy when $E_F=1$ eV.

\begin{figure}[htbp]
 \begin{center}
  \includegraphics[scale=0.6]{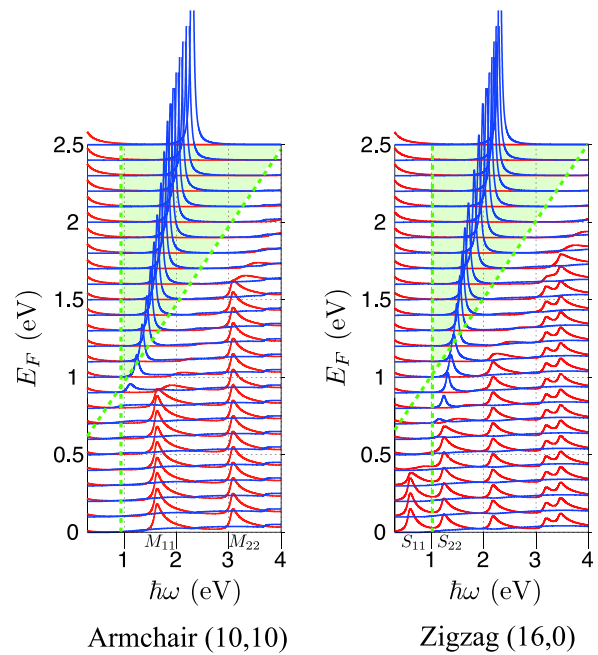}
 \end{center}
 \caption{{\bf Doping dependencies of absorption spectra.}
 Red (blue) spectra show the absorption of ${\bf E}_\parallel$ (${\bf
 E}_\perp$). 
 From the change in color of absorption peak, it is clear that the
 polarization dependence of the peak is reversed by doping. 
 Doping with $E_F \approx 1$ eV creates a transient region where 
 the polarization anisotropy starts to be reversed.
 The green dotted line shows the
 boundary of single particle excitation, and the green area
 ($\hbar \omega>2\hbar v/d_t$ and $\hbar\omega<2|E_F|-2\hbar v/d_t$ where $v$ is the
 Fermi velocity) shows the region where single particle excitation does
 not exist.~\cite{Sasaki2012b}
 }
 \label{fig:wf}
\end{figure}

Figure~\ref{fig:wf} shows the details of the $E_F$ dependence of the
absorption spectra of a $(10,10)$ armchair ($d_t=1.35$ nm) and $(16,0)$
zigzag CNTs ($d_t=1.25$ nm). 
There are several noticeable features should be mentioned.
Firstly, the plasmon peak starts to develop
when the $M_{11}$ ($S_{22}$) peak by ${\bf E}_\parallel$ starts to
disappear for the armchair (zigzag) CNT.
Secondly, 
the plasmon peak intensity and frequency increase as
increasing $E_F$.
Thirdly, 
the doping dependence of the plasmon frequency in the armchair CNT is
similar to that in the zigzag CNT.
This suggests that when CNTs are intentionally doped, 
they will eventually have a similar excitation structure regardless of
the chirality.
Finally, the plasmon peak is present in the dispersion region (or the
vicinity thereof) where single particle excitation is not allowed, 
indicating that a plasmon cannot collapse into individual electron-hole
pairs and the kinematic stability is guaranteed for the plasmon.

\section{Coulomb Interaction}\label{sec:exciton}

In this section, we examine how the Coulomb interaction affects the
results presented in the preceding sections.
Because the Coulomb interaction weakens at high doping or in a metallic
CNT due to the screening effect, we focus on a semiconducting CNT at low doping level. 
The results shown in this section are obtained by extending the existing
framework developed for calculating exciton of
an undoped CNT~\cite{Ando1997,Uryu2006} to a doped CNT. 
The details will be presented elsewhere.~\footnote{The results shown in
this section are dependent on an energy cutoff $E_c$ which limits the
electron or hole states being taken into account in evaluating the
physical quantities. We set $E_c=\gamma$.}

The Coulomb interaction changes the absorption spectrum through two
main effects: 
self-energy correction to the band-diagram (band renormalization) 
and formation of excitons.
First we show the band renormalization.

\subsection{Band Renormalization}

\begin{figure}[htbp]
 \begin{center}
  \includegraphics[scale=0.3]{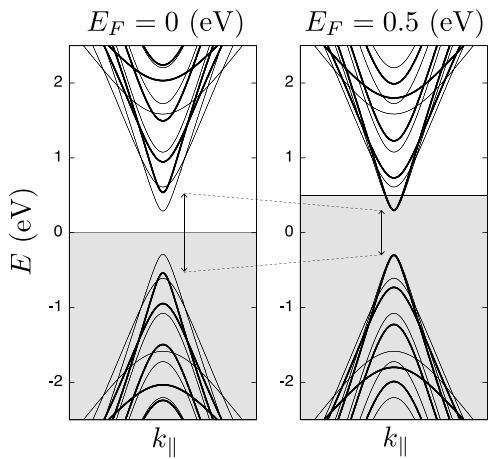}
 \end{center}
 \caption{{\bf Band renormalization of zigzag CNTs.}
 Thick (Thin) curved lines represent the renormalized (bare) band. 
 The self-energy of the lowest energy subbands is removed by modest
 doping ($E_F=0.5$ eV) so that the band gap decreases.
 }
 \label{fig:bgr}
\end{figure}

The thick curved lines in Fig.~\ref{fig:bgr} show the renormalized
band-diagram of a $(16,0)$ zigzag CNT, which is given by adding the
screened exchange energy (or self-energy) to the original (bare)
band-diagram denoted by the thin curved lines.
When $E_F=0$ eV, 
the self-energy makes the band gap increase significantly.
When $E_F=0.5$ eV, on the other hand,
the self-energy is modest;
the band gap is almost identical to that of the bare band.
This is due to that the electron-hole pairs within the conduction band
screen more effectively the Coulomb interaction than the inter-band
electron-hole pairs.~\cite{Sasaki2012b}
Note also that the self-energy for the states away from the Fermi
level does not vanish and this tends to blue-shift the plasmon peak.

\subsection{Absorption Spectra}

\begin{figure}[htbp]
 \begin{center}
  \includegraphics[scale=0.25]{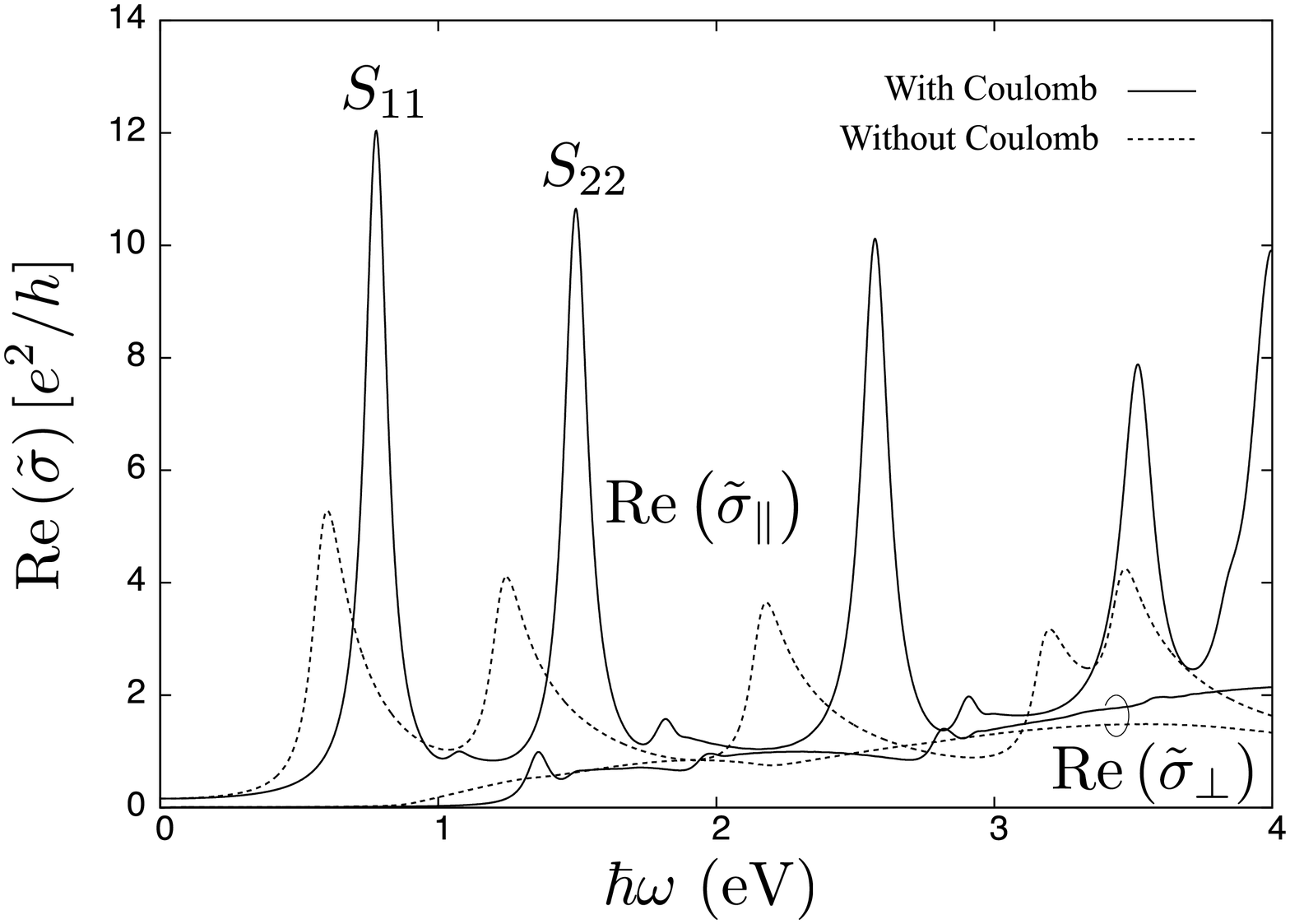}
 \end{center}
 \caption{{\bf Absorption spectra calculated with and without Coulomb interaction.}
 ${\rm Re}(\tilde{\sigma}_\parallel)$ and ${\rm Re}(\tilde{\sigma}_\perp)$ of (16,0) undoped CNTs are shown. 
 }
 \label{fig:abs_coul}
\end{figure}

The exciton formation together with the band renormalization changes 
the absorption spectrum significantly.
When $E_F=0$ eV, 
the absorption peaks of ${\bf E}_\parallel$ are governed by excitons as
shown in Fig.~\ref{fig:abs_coul}.
By comparing the result with the spectrum calculated without the Coulomb
interaction, the sizable enhancement of oscillator strength is seen for
each peak.
Meanwhile, 
the correction to the absorption spectrum of ${\bf E}_\perp$ is minor:
a small peak due to the exciton formation is observed in 
${\rm Re}(\tilde{\sigma}_\perp)$.
These results are consistent with Refs.~\onlinecite{Ando1997} and~\onlinecite{Uryu2006}.

\begin{figure}[htbp]
 \begin{center}
  \includegraphics[scale=0.4]{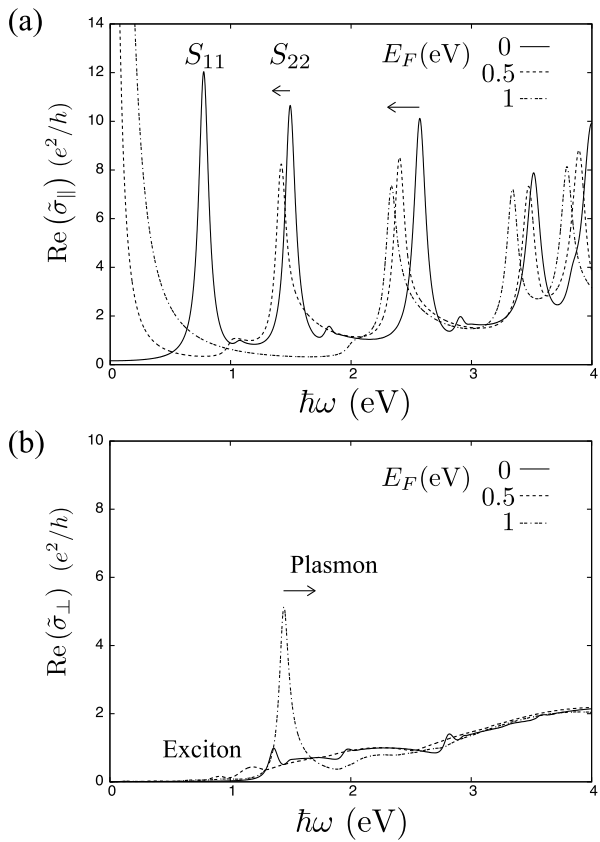}
 \end{center}
 \caption{{\bf Doping dependence of absorption spectra with Coulomb
 interaction.}
 (a) When $E_F=0.5$ eV, the $S_{11}$ peak is invisible due to the
 exclusion principle and the doping dependence of the $S_{22}$ peak
 exhibits a red-shift as indicated by the horizontal arrow.
 (b) The small exciton peak when $E_F=0$ eV is replaced by plasmon peak when $E_F=1$ eV.
 This transition may be observed when the exciton peak is observable.
 The arrow above the plasmon peak shows that the plasmon energy
 increases with increasing doping.}
 \label{fig:abs_coul_dope}
\end{figure}

Figure~\ref{fig:abs_coul_dope}(a) shows the behavior of the absorption
peaks of ${\bf E}_\parallel$ under doping. 
When $E_F=0.5$ eV, the $S_{11}$ peak disappears due to exclusion
principle.
The peak intensity of $S_{22}$ ($S_{33}$) is suppressed by doping.~\cite{Kazaoui1999}
Further increase of doping results
in that the $S_{22}$ and $S_{33}$ peaks exhibit red-shift due to the
band renormalization.
The red-shift of the peaks serves as a unique information of the
self-energy because the peak positions should not change when the
self-energy correction is not taken into account (see
Fig.~\ref{fig:wf}).
As shown in Fig.~\ref{fig:abs_coul_dope}(b),
the exciton peak of ${\bf E}_\perp$ disappears soon after $E_F$ reaches
the bottom of the first subband ($E_F=0.5$ eV), and the plasmon peak
develops when $E_F=1$ eV.
Such transition from exciton to plasmon may be observed when the
broadening of the exciton of an undoped CNT is sufficiently suppressed.~\cite{Uryu2006}

\section{Discussion}\label{sec:discussion}

We compare the present results with experimental ones. 
Kazaoui {\it et al.} found a broad peak in the absorption by thin
films of heavily doped single-walled CNTs.~\cite{Kazaoui1999}
The pristine films consist of semiconducting and metallic CNTs 
since $S_{11}$ (0.68 eV), $S_{22}$ (1.2 eV), and $M_{11}$ (1.8 eV) are
all observed.
The doping-induced peak appears 
when $S_{11}$, $S_{22}$, and $M_{11}$ disappear by doping, 
which is consistent with our results. 
It was found that the peak energy depends on doping level:
1.07 eV (1.3 eV) for CBr$_{0.15}$ (CCs$_{0.10}$), while the details
about the dependence was unknown.
Igarashi {\it et al.} clarified that using electrochemical doping 
the peak energy increased with increasing the doping.~\cite{Igarashi2015}
They showed further that semiconducting and metallic single-walled CNTs
cause independently the absorption peak at approximately 1 eV.
These are consistent with our results. 
However, the calculated peak energy is slightly above ($\sim 0.1$ eV)
the experimental result.
This discrepancy warrants further examination.
Petit {\it et al.} showed that doping thin films
with naphthalene-lithium did not cause the corresponding absorption
peak even though the doping level is high enough to make an absorption peak.~\cite{Petit1999a}
This suggests an interesting possibility that 
the surrounding of CNTs is modulated by the doping and that 
doping has an influence on the plasmon absorption (such as the peak energy
and intensity) through a mechanism beyond
the description by static dielectric constant.

When fully verifying the proposed theory, 
it is desirable to orient CNTs that are doped and separated into a single chirality.
In the past, the depolarization effect was experimentally verified by
absorption and Raman spectroscopy in which undoped CNTs are oriented by
stretching the organic films on which they are dispersed~\cite{ichida04} 
or by controlling magnetic effects~\cite{Islam2004} or CNT growth
processes.~\cite{Murakami2005}
Although experiments have already been performed on the doping
dependence of light absorption for CNTs with a single chirality,~\cite{Igarashi2015} 
there is no corresponding absorption measurements for oriented and doped CNTs.
Recently, He {\it et al.} developed a technology for aligning CNTs
spontaneously by improving vacuum filtration,~\cite{He2016}
and this approach can be used for the purpose.

If doped CNTs can be oriented,
they will provide an opportunity for searching for novel phenomena even
if they are not separated into a single chirality.
Because the anisotropy of light absorption is related to
the anisotropy of the electron-phonon interaction, 
there is a strong possibility that characteristic signals of doping will
be explored by the polarized Raman spectroscopy.~\cite{Kalbac2009}
For example, in doped metallic CNTs, phonon frequency hardening 
have been observed in the manner that depends on the phonon eigenvector.~\cite{farhat07}
A phenomenon similar to it should be observed also for semiconducting CNTs.

The idea of the polarization reversal of light absorption in doped CNTs can
also be applied to doped graphene nanoribbons, because it has been shown
that the optical selection rule of nanoribbons is similar to that of
CNTs.~\cite{Sasaki2011}
However, a modification of Eq.~(\ref{eq:loopcor}) caused by the edge is
needed for nanoribbons and the coupling of nanoribbons to the substrate
must be taken into account.

Since the length of a CNT is finite in the axial direction, 
there is also a depolarization effect in the axial
direction.~\cite{Zhang2013}
The optical selection rule of finite length CNTs is obtained by
extending the calculations on a nanoribbon. 
Indeed, due to the formation of a standing wave by the ends of a CNT, 
it can be proved that there is a wavelength shift of roughly the
reciprocal of the axial length, which can explain why the plasmon peak
is formed in the terahertz region of ${\bf E}_\parallel$.~\cite{Zhang2013}

Here, we mention a subject closely related to the optical properties of
doped CNTs, that is, quantum wells.
The band-diagram of a CNT bears a similarity to that of a quantum
well and the concepts such as depolarization and exciton effects have
been used to understand the optical properties of quantum wells.
The term ``intersubband transitions''
is commonly used to describe only the transitions within the conduction band
of quantum wells.~\cite{Liu2000}
This is a reasonable assumption when the width of doped quantum wells is
approximately 10 nm or longer.
For CNTs with diameter of the order of 1 nm, however,
the inter-band transitions have very important effects on absorption spectrum
for both undoped and doped cases.
Note also that the pseudospin selection rule
is a fundamental new point of CNTs, not seen in quantum wells.

Two degrees of freedom of the light polarization are utilized in modern
optical transmission technology to double the amount of information
transmitted simultaneously.
For example, a light is propagated by associating its parallel
polarization with pictorial information and perpendicular polarization
with sound. 
Nano-scale materials that respond differently depending on polarization
direction are advantageous for information manipulation in highly refined
structures where light propagates, for example as an extremely thin
Polaroid film.
The fact that the polarization direction of light transmitted through
CNTs can be rotated by 90 degrees simply by doping implies the possibility
of performing further information manipulation by electric means.
From the viewpoint of condensed matter physics, 
the doping-induced change in the phase of the excited states from
excitons to plasmons is an intriguing topic.
Our conclusion is thus to stimulate both fundamental research on
CNTs and application research related to optical devices.

\section*{Acknowledgments}

The authors thank K. Yanagi and J. Kono for discussions.

\appendix

\section{Method}

We employed a tight-binding model with the hopping integral 
$\gamma=2.6$ eV and atomic distance $a_{cc}=1.42$ \AA, to calculate
the band diagram and wavefunction of the $\pi$ electrons in CNTs.~\cite{saito98book}
As a function of wavevector ${\bf k}=(k_x,k_y)$,
the model Hamiltonian is written in the form of a $2\times 2$ matrix:
\begin{align}
 H({\bf k}) = -\gamma
 \begin{pmatrix}
  0 & f({\bf k}) \cr 
  f({\bf k})^* & 0
 \end{pmatrix}.
 \label{app:H}
\end{align}
The off diagonal element $f({\bf k})$ is a complex number given by 
\begin{align}
 f({\bf k}) = e^{ik_y a_{cc}} + 2e^{-i\frac{k_y a_{cc}}{2}}
 \cos\left(\frac{\sqrt{3}k_x a_{cc}}{2}\right),
 \label{app:f}
\end{align}
and $f({\bf k})^*$ denotes the complex conjugate of $f({\bf k})$.
By expressing the energy eigenvalue equation 
$H({\bf k})|\phi^s_{\bf k} \rangle= \varepsilon^s_{\bf k}|\phi^s_{\bf k} \rangle$
in terms of the magnitude and phase of $f({\bf k})$ as 
$f({\bf k})=|f({\bf k})|e^{-i\Theta({\bf k})}$, 
we obtain the energy eigenvalue and Bloch wavefunction as
$\varepsilon^s_{\bf k}=-s \gamma|f({\bf k})|$ and
\begin{align}
 |\phi^s_{\bf k} \rangle = \frac{1}{\sqrt{2}}
 \begin{pmatrix}
  e^{-i\Theta({\bf k})} \cr s
 \end{pmatrix},
 \label{eq:wf}
\end{align}
respectively.
The band index $s=+1$ ($s=-1$) corresponds to the valence (conduction)
band. 
The low energy band diagram near the charge neutrally point
$\varepsilon_{\bf k}^s\sim 0$ is given by a pair of double cones (known
as the Dirac cones).

Because the interaction between the electron and light $\delta H$ is
given by the minimal substitution, $k_i \to k_i - \frac{e}{\hbar}A_i$, 
the electric currents defined from $\delta H = J_i A_i$ are 
\begin{align}
 J_i = - \frac{e}{\hbar} \frac{\partial H({\bf k})}{\partial k_i} \ \ 
 (i=x,y).
 \label{app:j}
\end{align}
Putting Eqs.~(\ref{app:H}) and (\ref{app:f}) into Eq.~(\ref{app:j}) we have
the following expressions of the current operators,
\begin{widetext}
\begin{align}
 & J_x = -ev \frac{2}{\sqrt{3}} \sin\left(\frac{\sqrt{3}k_x a_{cc}}{2}\right)
 \begin{pmatrix}
  0 & e^{-i\frac{k_y a_{cc}}{2}} \cr
  e^{+i\frac{k_y a_{cc}}{2}} & 0
 \end{pmatrix},
 \label{eq:Jx}
 \\
 & J_y = -e v \frac{2}{3} 
 \begin{pmatrix}
  0 & -i \left( e^{ik_y a_{cc}} -e^{-i\frac{k_y a_{cc}}{2}}
  \cos\left(\frac{\sqrt{3}k_x a_{cc}}{2}\right) \right) \cr
  i \left( e^{-ik_y a_{cc}} -e^{i\frac{k_y a_{cc}}{2}}
  \cos\left(\frac{\sqrt{3}k_x a_{cc}}{2}\right) \right) & 0
 \end{pmatrix},
 \label{eq:Jy}
\end{align}
\end{widetext}
where $v\equiv 3\gamma a_{cc}/2\hbar$ is the Fermi velocity of graphene.

The wavevectors are quantized by the periodic boundary condition around
and along the tube's axis.~\cite{Malic2006}
For the case of armchair $(n,n)$ CNTs, 
the quantized wavevectors are specified by two integers $m$ and $j$ as 
\begin{align}
 & k_y \frac{3a_{cc}}{2} n = m \pi, \ \ (m=-n+1,\ldots, n), \\
 & k_x \frac{\sqrt{3}a_{cc}}{2} 2L = j\pi, \ \ (j=-L,\ldots,L).
\end{align}
Note that $2\sqrt{3}a_{cc}L$ is the CNT length, $3a_{cc}n/\pi$ is the
diameter, and the surface area of a CNT ($S$) is $3a_{cc} n \times 2\sqrt{3}a_{cc}L$.
As a result of the longer axial length than the diameter ($L\gg n$), the
band diagram of a CNT is well described by the cross sections of the
Dirac cone (see Fig.~\ref{fig:Absorption_szig}(b)).
We also note that the effect of orbital hybridization between $\pi$ and
$\sigma$ due to the curvature of the azimuthal direction is negligible
in this study.
For the case of zigzag $(n,0)$ CNTs, 
the quantized wavevectors are specified by two integers $m$ and $j$ as 
\begin{align}
 & k_y \frac{3a_{cc}}{2} 2L = j \pi, \ \ (j=-L,\ldots, L), \\
 & k_x \frac{\sqrt{3}a_{cc}}{2} n = m\pi, \ \ (m=-n+1,\ldots,n).
\end{align}
Note that $6a_{cc}L$ is the CNT length, $\sqrt{3}a_{cc}n/\pi$ is the
diameter, and the surface area of a CNT is the same as $(n,n)$ CNTs.

We calculated the dynamical conductivity in the framework of the linear
response theory, 
\begin{align}
 & \sigma_{\Delta m}(\omega,E_F) \equiv g_{\rm spin}
 \frac{\hbar}{iS} \sum_{s',s}\sum_{m,j} \nn \\
 &
 \frac{\left(f^{s'}_{m+\Delta m,j}(E_F)-f^s_{m,j}(E_F)\right)
 \left| \langle \phi^{s'}_{m+\Delta m,j} | J_i | \phi^{s}_{m,j} \rangle
 \right|^2}{\left(\varepsilon^{s'}_{m+\Delta m,j}-\varepsilon^s_{m,j}\right)
 \left(\varepsilon^{s'}_{m+\Delta m,j}-\varepsilon^s_{m,j}+ \hbar \omega +i\delta
 \right)},
\end{align}
where $g_{\rm spin}=2$ is the spin degeneracy, 
$f^s_{\bf k}(E_F)=1/(e^{(\varepsilon^s_{\bf k}-E_F)/kT} + 1)$ is the
Fermi distribution function at room temperature ($kT=1/38.6$ eV),
and $\delta$ ($=\hbar/\tau$) is inversely proportional to the
relaxation time of an excited electron. 
We fix $\delta=50$ meV ($\tau\approx 13$ fs) in all calculations.
For an armchair CNT, 
the current operator $J_x$ ($J_y$) couples to
${\bf E}_\parallel$ (${\bf E}_\perp$).
Due to the momentum selection rule $\Delta m=0$ for ${\bf E}_\parallel$,
the absorption $J_x E_x$ is the product of 
\begin{align}
 \sigma_0
 \label{app:para}
\end{align}
and $E_x^2$, while for ${\bf E}_\perp$, the absorption $J_y E_y$ is the
product of 
\begin{align}
 \frac{1}{2}
 \left(
 \frac{\sigma_{+1}}{1-\frac{\sigma_{+1}}{i\omega \epsilon d_t}}
 +\frac{\sigma_{-1}}{1-\frac{\sigma_{-1}}{i\omega \epsilon d_t}}
 \right)
 \label{app:perp}
\end{align}
and $E_y^2$.
The factor one-half in Eq.~(\ref{app:perp}) originates from the field
decomposition of $E_y$ into the pair $\Delta m = \pm 1$.
Equations~(\ref{app:para}) and (\ref{app:perp}) are the exact definition
of the absorption plotted in the text as $\sigma_\parallel$ and
$\tilde{\sigma}_\perp$, respectively.
Note also that $\sigma_{+1}=\sigma_{-1}(\equiv \sigma_\perp)$ holds in
the absence of the Aharonov-Bohm flux along the tubule axis.

It is instructive to evaluate the matrix element to show that only the
forward scattering is allowed by the selection rule of pseudospin.
We take zigzag CNTs and focus on the transitions between band edges
($k_y=0$). 
The matrix element of $J_x$ is known from Eqs.~(\ref{eq:wf}) and (\ref{eq:Jx}) as 
\begin{align}
 \langle \phi^{s'}_{m'} | J_x | \phi^{s}_{m} \rangle 
 \propto 1 + ss' e^{i(\Theta'+\Theta)}.
\end{align}
Thus, for the inter-band transitions ($ss'=-1$), the transitions
satisfying $\Theta'+\Theta=\pi$, which are the forward scattering, have
the largest matrix element squared.
For the intra-band transitions ($ss'=1$), the transitions satisfying
$\Theta'+\Theta=0$, which are the forward scattering too, are allowed.  
The intra-band backward scattering satisfies $\Theta'+\Theta=\pi$ and
has vanishing matrix element.~\cite{Sasaki2012b}

\begin{figure}[htbp]
 \begin{center}
  \includegraphics[scale=0.3]{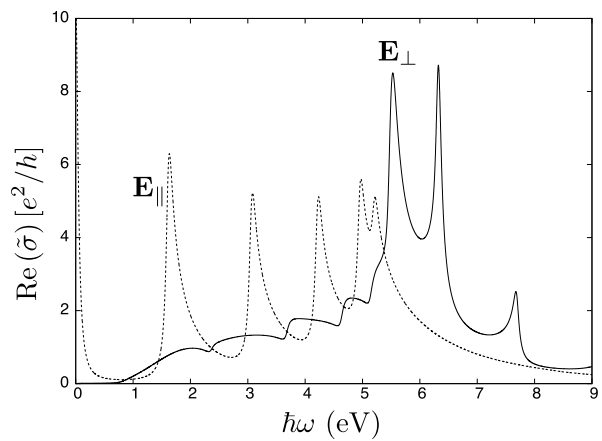}
 \end{center}
 \caption{{\bf Absorption spectra of (10,10) armchair CNTs.}
 The depolarization effect of ${\bf E}_\perp$ suppresses a peak to
 develop in ${\rm Re}(\tilde{\sigma}_\perp)$ at low energy ($\hbar \omega
 \approx 1$ eV).
 However, ${\rm Re}(\tilde{\sigma}_\perp)$ has the peaks at high energy around 6
 eV. 
 }
 \label{fig:abs_highene}
\end{figure}

The polarization
characteristics of the absorption spectrum has been investigated for
incident light energies up to 6 eV.~\cite{Murakami2005}
The absorption peaks observed at 4.5 and 5.25 eV were found to exhibit
different polarization dependencies.
The behavior is also reproduced by our calculation, that is, 
peaks caused by ${\bf E}_{\parallel}$ and ${\bf E}_{\perp}$ appear
approximately at 5 and 6 eV, respectively, as shown in Fig.~\ref{fig:abs_highene}.
The discrepancy between experiment and calculation may be attributed to
the fact that we have ignored the overlap 
between the wavefunctions at nearest neighbor $\pi$ electrons
giving asymmetry in the conduction and valence bands.~\cite{saito98book}
The qualitative agreement suggests the
correctness of the optical matrix elements used to evaluate the
dynamical conductivity. 
Note that the peak structure of ${\rm Re}(\tilde{\sigma}_\perp)$ at about 6
eV may be regarded as a plasmon resonance because 
${\rm Im}(\sigma_\perp)$ has a peak structure at the energy and 
${\rm Re}(\varepsilon_\perp)$ approaches zero.
Note also that a realistic doping level does not change these
high-energy peaks and that the problem can be discussed simply in terms
of the photo-electron interaction and the density of the electronic
state.~\cite{Shyu1999}

\bibliographystyle{apsrev4-1}
%

%

\end{document}